\documentclass[10pt,article,twocolumn,superscriptaddress,showpacs,floatfix]{revtex4-2}
\usepackage{graphics}
\usepackage{graphicx}
\usepackage{bm}
\usepackage{amsmath}
\usepackage{amsfonts}
\usepackage{amssymb}
\usepackage{latexsym}
\usepackage{color}

\usepackage{hyperref}

\begin{document}

\title{Mirror-mediated ultralong-range atomic dipole-dipole interactions}
\author{Nicholas Furtak-Wells} 
\affiliation{The School of Physics and Astronomy, University of Leeds, Leeds LS2 9JT, United Kingdom}
\author{Benjamin Dawson}
\affiliation{The School of Physics and Astronomy, University of Leeds, Leeds LS2 9JT, United Kingdom}
\affiliation{School of Chemical and Process Engineering, University of Leeds, Leeds LS2 9JT, United Kingdom}
\author{Thomas Mann}
\affiliation{School of Chemical and Process Engineering, University of Leeds, Leeds LS2 9JT, United Kingdom}
\author{Gin Jose} 
\affiliation{School of Chemical and Process Engineering, University of Leeds, Leeds LS2 9JT, United Kingdom}
\author{Almut Beige}
\affiliation{The School of Physics and Astronomy, University of Leeds, Leeds LS2 9JT, United Kingdom}
\date{\today}

\begin{abstract}
In three dimensions, dipole-dipole interactions which alter atomic level shifts and spontaneous decay rates only persist over distances comparable to the wavelength of the emitted light. In this paper we show that it is possible to significantly extend the range of these interactions with the help of a partially transparent asymmetric mirror interface. Suppose two two-level atoms are placed on opposite sides of the interface, each at the position of the mirror image of the other. In this case, their emitted light interferes almost exactly as it would when the atoms are right next to each other. Hence their dipole-dipole interaction assumes an additional maximum, even when the actual distance of the atoms is several orders of magnitude larger than the transition wavelength. Although the resulting ultralong-range interactions are in general relatively weak, we expect them to find applications in quantum technology, like non-invasive quantum sensing.
\end{abstract}

\maketitle

\section{Introduction} 

\begin{figure*}[t] 
	\centering
	\includegraphics[width=0.9 \textwidth]{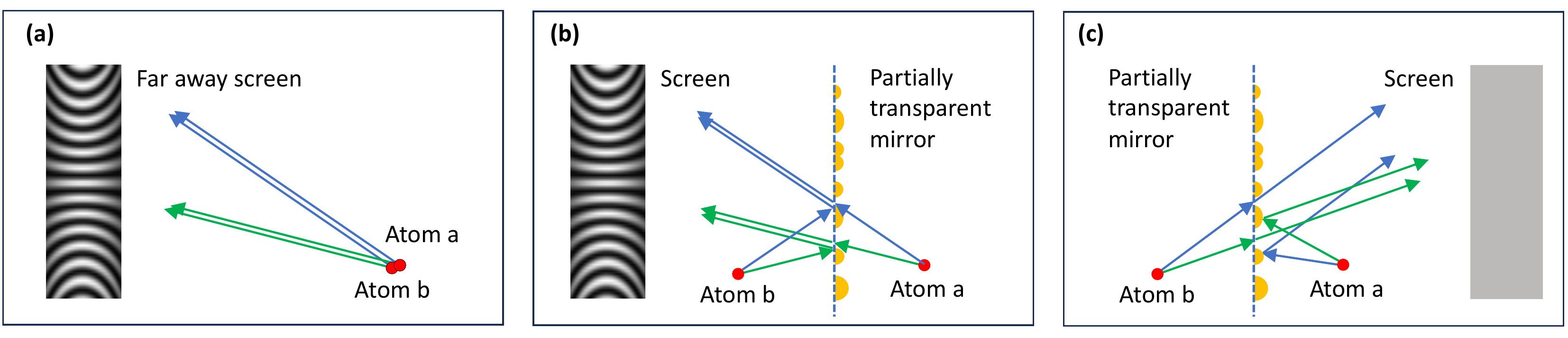} 
\caption{(a) Schematic view of the two-atom double-slit experiment by Eichmann {\em et al.}~\cite{Wineland}. The emitted light interferes either constructively or destructively when arriving at a far away-screen. Which one applies depends only on the collective state of the atoms, their distance and the direction of propagation of the emitted light. As first pointed out by Dicke \cite{Dicke}, for relatively small distances between the atoms, the interference results in dipole-dipole interactions: spontaneous decay rates change (sub- and superradiance) and atomic level shifts occur. (b) Schematic view of two atoms on opposite sides of a partially transparent asymmetric mirror which is smooth on one side but rough on the other. Such a mirror can be realised by placing tiny metallic islands (represented by yellow semicircles) onto a glass surface, while leaving small gaps between them. If the mirror interface is smooth on the left hand side, the transmitted light coming from atom $a$ and the reflected light coming from atom $b$ interfere exactly as in Fig.~1(a) and an analogous interference pattern emerges, if a screen is placed on the left hand side of the setup. (c) On the right hand side of the mirror interface, the reflected light travels in different (essentially random) directions. Hence the reflected light coming from atom $a$ and the transmitted light coming from atom $b$ no longer travel exactly the same distance before arriving at the same point on a far-away screen and the interference pattern disappears.}
	\label{fig1}
\end{figure*}

In 1982, Scully and Dr\"uhl \cite{Scully} proposed a double-slit experiment in which the slits are two two-level atoms. As illustrated in Fig.~\ref{fig1}(a), the atoms are kept at a constant distance, are continuously driven by laser light and emit photons at a constant rate. When their distance is comparable to the wavelength of the emitted light, an interference pattern forms on a far-away screen. Averaged over many photons, this pattern very closely resembles the interference pattern of classical double-slit experiments. It only disappears when information about the origin of each photon becomes available \cite{Wineland}. As in classical two-slit interference experiments, the distance between the intensity minima and maxima depends on the distance between the atoms \cite{Schoen}. 

When this two-atom double-slit experiment was first performed by Eichmann {\em et al.}~in 1993 \cite{Wineland}, it raised many questions, like, how can spontaneously emitted photons interfere \cite{Pachos}. A closer look at the experiment shows that it is best to think of the atoms as continuously radiating dipole antennae \cite{NP}. Both atoms constantly transfer energy into the surrounding free radiation field which only manifests itself as ``individual photons" upon detection \cite{Hegerfeldt,Stokes}. When an individual photon is registered on a photographic plate, it contains in general energy from {\em both} atoms. Depending on the direction of emission, radiation either interferes constructively or destructively, thereby resulting in a spatial dependence of the intensity of the emitted light. Moreover, interference effects result in a spatial dependence of first and second order photon correlations \cite{BeHe,Antoine,Schmidt-Kaler}. By now, the interference of light from distant atoms is relatively well understood and has  already found applications in distributed quantum computing \cite{Kok,Lim,Monroe,Hanson,Lukas}, in designing mirrors with unusual properties \cite{Cheng}, and in quantum sensing \cite{Faccio}. 

Different from the classical case, interference in the two-atom double-slit experiment depends on the internal state of the slits, since different entangled atomic states radiate light in different preferred directions \cite{new2,Blatt,new3}. Suppose two atoms are right next to each other and share a single energy quantum. If the atoms in Fig.~\ref{fig1}(a) are in their maximally-entangled symmetric state, all of the emitted light interferes constructively. The atomic coupling to the free radiation field is collectively enhanced and a photon is emitted at twice the usual rate. However, if the atoms are in their anti-symmetric state, their efforts to transfer energy into the free radiation field cancel each other out. The spontaneous decay rate of the antisymmetric state therefore tends to zero. At finite distances between the atoms which are of the order of the wavelength of the emitted light, similar alterations of spontaneous decay rates occur. These are synonymous with Dicke sub- and superrandiance \cite{Dicke} and indicate the presence of atomic dipole-dipole interactions \cite{Agarwal,Haroche,new1,Ficek}. A possible approach to atomic ultralong-range interactions is therefore the recreation of the interference effects of the original two-atom double-slit experiment for large atomic distances. Taking this into account, this paper predicts mirror-mediated, targeted ultralong-range dipole-dipole interactions which can persist over distances that are many orders of magnitude larger than the wavelength of the emitted light. 

Suppose two two-level atoms, $a$ and $b$, are separated by a two-sided partially transparent mirror, i.e.~a surface with finite reflection and transmission rates $r_i$ and $t_i$ $(i=a,b)$ \cite{LSPRA,Abeer}, and the position of each atom coincides with the position of the mirror image of the atom on the opposite side, as illustrated in Fig.~\ref{fig1}(b). Comparing Figs.~\ref{fig1}(a) and (b) and viewing the atoms as radiating dipole antennae, we see that\textemdash for half of the emitted light\textemdash the two paths from a source to a certain point on the far-away screen are always of the same length. The resulting interference pattern is therefore the same as in the above described two-atom double-slit experiment \cite{Scully}, apart from a reduction in visibility. Since atomic dipole-dipole interactions are the result of interference effects and the interference of spontaneously emitted photons is the same in Figs.~\ref{fig1}(a) and (b), the above discussion suggests an additional maximum of the dipole-dipole interaction between two atoms on opposite sides of a partially transparent mirror. As we shall see below, the strength of this ultralong-range interaction does {\em not} depend on the actual distance of the atoms but on the distance between atom $a$ and the mirror image of atom $b$.

Since light coming from atom $a$ and light coming from atom $b$ travels the same distance before arriving at the the same point on the screen, it only depends on the initial state of the atoms whether the resulting interfere is constructive or destructive. Consequently, as we shall see below, certain collective atomic states decay more rapidly, while the decay of other collective atomic states gets delayed. This effect manifests itself in an alteration of the spontaneous decay rates of the atoms. For {\em symmetric} mirrors, which reflect light such that the angle of incidence always equals the angle of reflection, it can be shown that the predicted mirror-mediated atomic interactions scale as $ r_a^* t_b + t_a^* r_b$. Unfortunately, we know from classical optics that symmetric mirrors only conserve the energy of any incoming wave packets when 
\cite{LSPRA,Abeer}
\begin{eqnarray} \label{SL}
|r_i|^2 +|t_i|^2 = 1 \, , ~~ r_a^* t_b + t_a^* r_b = 0 \, .
\end{eqnarray}
This means, symmetric mirrors cannot alter the spontaneous decay rates of atoms on opposite sides of a partially transparent interface. Generating remote dipole-dipole interactions therefore requires the presence of an asymmetric mirror.

One way of realising an {\em asymmetric} mirror is to vary the surface roughness on both sides of the interface, as illustrated in~Figs.~\ref{fig1}(b) and (c). For simplicity, we assume in the following that the partially transparent mirror is smooth on one side but uneven on the other. Such a mirror is obtained, for example, after placing tiny metallic droplets onto a glass surface with some space (tiny holes) between them. On the side of the glass, the surface of the metal is smooth and light is reflected as it would be in case of a symmetric mirror (cf.~Figs.~\ref{fig1}(b)), while the droplets on the other side reflect light essentially in random directions (cf.~Figs.~\ref{fig1}(c)). Light arriving at the holes, however, travels through without changing direction, as long as the metal islands and the holes are much smaller than the optical wavelength and relatively evenly distributed. When comparing Figs.~\ref{fig1}(b) and (c), we see that the situation is very different in both cases. Now, light which is emitted into the same direction travels a different distance when coming from atom $a$ and when coming from atom $b$. As a result, all light emitted to the right side contributes equally to the spontaneous decay rates of the atoms. As we shall see below, the predicted interaction therefore scales as $ t_a^* r_b$ in this case which is in general non-zero.

Deriving the quantum optical master equations for the experimental setup in Fig.~\ref{fig1}, while assuming that the actual distance of the atoms is relatively large, shows that their spontaneous decay rates are formally the same as in the case of two nearby atoms with free space dipole-dipole interactions as long as atom $a$ is close to the mirror image of atom $b$. However, the actual distance of the atoms can now be several order of magnitude larger than the wavelength of the emitted light. This is not surprising, since dipole-dipole interactions \cite{Agarwal,Haroche,new1,Ficek} are mediated by photons and photons can travel relatively large distances one the timescale of the fluorescence lifetimes. For example, the effective length of a spontaneously emitted photon from a single trapped ion in free space easily exceeds one meter which makes an interaction range of mirror-mediated dipole-dipole interactions of the order of several millimeters plausible. As mentioned already above, the main obstacle to generating ultralong-range atomic interactions is our ability to control the interference of the emitted light without also having to control its direction of propagation.

Atomic dipole-dipole interactions have already been studied in different environments but so far they have always been relatively short-range \cite{Klim2,Klim3}. The only exception are atoms which couple to the common field mode inside an optical cavity \cite{cavity}. Theoretical and experimental studies usually consider atom-mirror interactions \cite{Drexhage,Barnes,Eschner,Hoi}, interactions between atoms on the same side of an interface \cite{Palacino,Meystre,Zhou3}, atoms separated by negative index metamaterials and other thin films \cite{Pendry,2,1,Michael,3,6,5,4,7} and atoms near one-dimensional nanofibers and wave guides \cite{Vidal,Rauschenbeutel,Kimble,Rolston}. Although the mirror-mediated atomic interactions which we predict in this paper are weaker than the standard dipole-dipole interactions of nearby atoms, they are expected to find applications, for example, in non-invasive quantum sensing based on fluorescence lifetime measurements.

\section{Results}

\subsection{Local atom-field interactions}

In free space, the complex electric field observable $\boldsymbol{\cal E}(\boldsymbol{r})$ at position $\boldsymbol{r}$ can be written as a superposition of local contributions $\boldsymbol{\cal E}_{{\boldsymbol s} \lambda} (\boldsymbol{r})$ of travelling waves with polarisations $\lambda = 1,2$ and directions of propagation $\boldsymbol{s}$,
\begin{eqnarray} \label{eq:observables}
\boldsymbol{\cal E} (\boldsymbol{r}) = \sum_{\lambda = 1,2} \int_{\cal S} {\rm d} \boldsymbol{s} \, \boldsymbol{\cal E}_{\boldsymbol{s} \lambda} (\boldsymbol{r}) \, .
\end{eqnarray}
Here ${\cal S}$ denotes the set of all three-dimensional unit vectors and the operator $\boldsymbol{\cal E}_{\boldsymbol{s} \lambda} (\boldsymbol{r})$ creates local photons with wave vectors ${\boldsymbol k} = k {\boldsymbol s}$, normalised polarisation vectors $\boldsymbol {e}_{{\boldsymbol s} \lambda}$ with $\boldsymbol{e}_{\boldsymbol{s} 1} \cdot \boldsymbol{e}_{\boldsymbol{s} 2} =  \boldsymbol{e}_{\boldsymbol{s} \lambda} \cdot \boldsymbol{s} = 0$, and bosonic creation operators $a^\dagger_{{\boldsymbol k} \lambda} $. Using this notation, $\boldsymbol{\cal E}_{\boldsymbol{s} \lambda} (\boldsymbol{r})$ can be written as \cite{Bennett}
\begin{eqnarray} \label{eq:rho3xxxxx}
\boldsymbol{\cal E}_{\boldsymbol{s} \lambda} (\boldsymbol{r}) 
= - {\rm i} \int_0^\infty {\rm d} k \, k^2 \, \left( {\hbar c k \over 16 \pi^3 \varepsilon} \right)^{1/2} {\rm e}^{-{\rm i} {\boldsymbol k} \cdot {\boldsymbol r}} \, a^\dagger_{{\boldsymbol k} \lambda} \,  \boldsymbol{e}_{\boldsymbol{s} \lambda} \, . 
\end{eqnarray}
Suppose $ |0_{\rm F} \rangle$ and $U_{\rm F}(t,0) $ denote the vacuum state and the time evolution operator of the free field Hamiltonian $H_{\rm F}$, respectively. Then 
\begin{eqnarray} \label{eq:dyn}
U_{\rm F} (t,0) \, \boldsymbol{\cal E}_{\boldsymbol{s} \lambda} (\boldsymbol{r}) \, |0_{\rm F} \rangle = \boldsymbol{\cal E}_{\boldsymbol{s} \lambda} (\boldsymbol{r} + \boldsymbol{s} ct) \, |0_{\rm F} \rangle \, ,
\end{eqnarray}
since a local field excitation with a well defined direction of propagation $\boldsymbol{s}$ simply travels at the speed of light $c$ in a straight line away from its source \cite{Jake,Daniel}. If created at an initial time $t=0$ at position $\boldsymbol{r}$, it will be found at position $\boldsymbol{r} + \boldsymbol{s} ct$ at some later time $t$.

Next we assume that a partially transparent asymmetric metasurface is placed in the $x=0$ plane, as illustrated in Figs.~\ref{fig1}(b) and (c). Suppose this mirror is obtained by placing a thin metallic film with tiny holes which are much smaller than the wavelength of the emitted light onto a glass surface. In this case, the local field excitations which meet the gaps are transmitted and evolve exactly as they would in free space (i.e.~as in Eq.~(\ref{eq:dyn})). However, 
light which does not meet a hole, is reflected and evolves such that
\begin{eqnarray}\label{Efieldinterfacenew2}
U_{\rm F} (t,0) \, \boldsymbol{\cal E}_{\boldsymbol{s} \lambda} (\boldsymbol{r}_b) \, |0_{\rm F} \rangle 
= \boldsymbol{\cal E}_{\tilde{\boldsymbol{s}} \lambda} (\tilde{\boldsymbol{r}}_b + \tilde{\boldsymbol{s}} ct) \, |0_{\rm F} \rangle \, ,
\end{eqnarray}
if it has been created at the position $\boldsymbol{r}_b$ of atom $b$ at $t=0$ and if the mirror surface is smooth on the left. The tilde indicates that a minus sign has been added to the $x$ component of the respective vector, thereby ensuring for example that electric field vectors are always orthogonal to their direction of propagation. Similarly, for a metasurface which is rough on the right, Eq.~(\ref{eq:dyn}) changes into
\begin{eqnarray}\label{Efieldinterfacenew22}
U_{\rm F} (t,0) \, \boldsymbol{\cal E}_{\boldsymbol{s} \lambda} (\boldsymbol{r}_a) \, |0_{\rm F} \rangle 
= \boldsymbol{\cal E}_{{\boldsymbol S}({\boldsymbol s}) \lambda} ({\boldsymbol R}_a({\boldsymbol s},t)) \, |0_{\rm F} \rangle 
\end{eqnarray}
for reflected light originating from atom $a$ at position $\boldsymbol{r}_a$ at $t=0$. Here ${\boldsymbol S}({\boldsymbol s})$ and ${\boldsymbol R}_a({\boldsymbol s},t)$ denote the direction of propagation and the position of the respective $({\boldsymbol s}, \lambda)$ field excitation at time $t$. The exact values of these two variables do not need to be known. All we need to take into account is that the surface roughness stops transmitted and reflected light from interfering efficiently on the right hand side of the mirror interface. The only assumption we make in the Methods section for simplicity is that the ${\boldsymbol S}({\boldsymbol s})$ vectors cover the right hand side of the $x=0$ plane relatively evenly.

In the following, we denote the electron charge and the complex dipole moment of atom $i$ with ground state $|1\rangle_i $ and excited state $|2 \rangle_i $ by $e$ and $\boldsymbol{D}_{12}^{(i)}$, respectively. Then, within the dipole and the rotating wave approximation, the interaction Hamiltonian between the atoms and the surrounding free radiation field can be written as \cite{Agarwal,Haroche,new1,Ficek}
\begin{eqnarray} \label{eq:Hdef2}
H_{\rm AF} = e \sum_{i=a,b} \boldsymbol{D}_{12}^{(i)} \sigma_i^- \cdot \boldsymbol{\cal E} (\boldsymbol{r}_i) + {\rm H.c.}  
\end{eqnarray}
with $\sigma_i^- = |1 \rangle_{ii} \langle 2|$ denoting the lowering operator of atom $i$. In Methods, we analyse the dynamics generated by this Hamiltonian using second order perturbation theory. As we shall see below, as long as we know how the atomic operators $\sigma_i^-$ and the local electric field observable $\boldsymbol{\cal E}_{\boldsymbol{s} \lambda} (\boldsymbol{r}_i)$ evolve in the absence of atom-field interactions, the dynamics of the two two-level atoms in Fig.~\ref{fig1} can be analysed in a relatively straightforward way. 

\subsection{Dynamics of atomic states}

Quantum optical master equations describe the dynamics of atomic density matrices $\rho_{\rm A}(t)$ on a coarse grained time scale $\Delta t$ which is much larger than their inverse transition frequency $1/\omega_0$ but also much smaller than their atomic lifetime $1/\Gamma_{\rm free}$ \cite{Hegerfeldt,Stokes}. To derive the master equations for the experimental setup in Fig.~\ref{fig1}, we assume in the following that the free radiation field is initially in its vacuum state $|0_{\rm F} \rangle$, evolve atoms and field for a time $\Delta t$ with their Hamiltonian $H$ in Eq.~(\ref{H}) and follow these dynamics with a measurement as to whether or not a photon has been emitted. Proceeding as described in Methods, one can then show that the time derivative of the atomic density matrix $\rho_{\rm A}$ equals
\begin{eqnarray} \label{eq:rho30}
\dot{\rho}_{\rm A} = - {{\rm i} \over \hbar} \left( H_{\rm cond} \rho_{\rm A} - \rho_{\rm A} H_{\rm cond}^\dagger \right) + {\cal L} (\rho_{\rm A}) ~~
\end{eqnarray}
to a very good approximation. The reset operator ${\cal L} (\rho_{\rm A})$ and the non-Hermitian Hamiltonian $H_{\rm cond}$ in this equation can be used to analyse the dynamics of the two two-level atoms in a time interval $(0,\Delta t)$ under the condition of a photon emission and no emission, respectively. 

\begin{figure*}[t] 
	\centering
	\includegraphics[width=0.95 \textwidth]{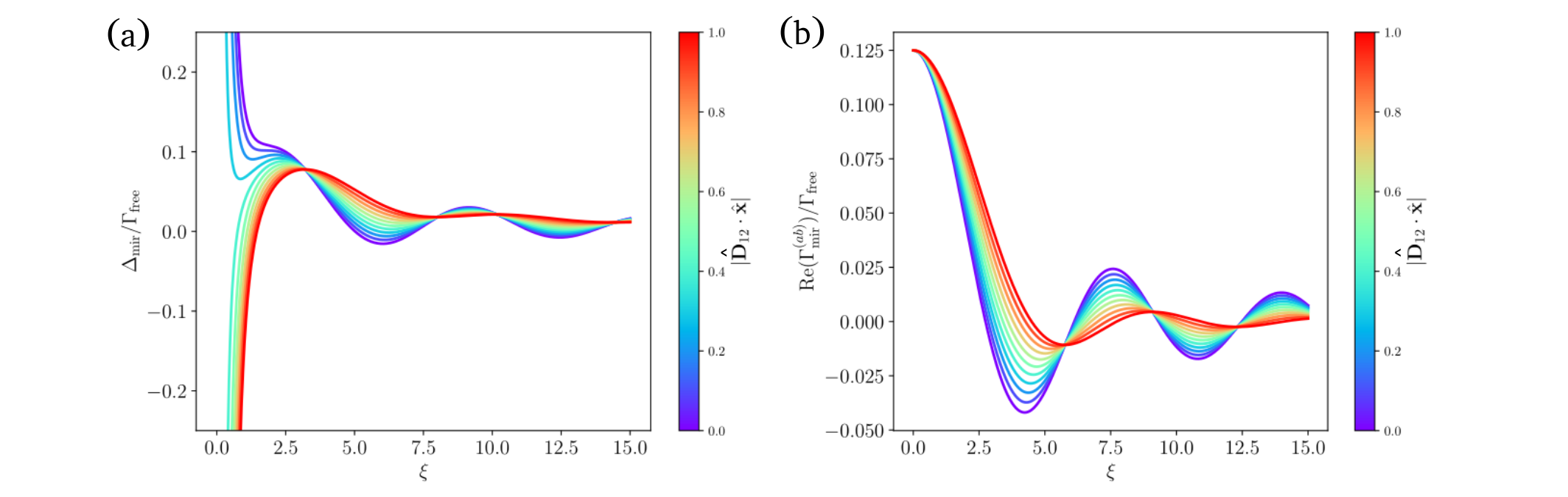} 
\caption{(a) The imaginary part $\Delta_{\rm mir} = {\rm Im}(\Gamma_{\rm mir}^{(ab)})$ which is responsible for the level shifts of certain collective atomic states as a function of the distance $\xi$ between atom $a$ and the mirror image of atom $b$ for different orientations of the atomic dipole moment vectors ${\boldsymbol D}^{(a)}_{12}$ and ${\boldsymbol D}^{(b)}_{12}$ for $t_a r_b = 0.5$.  For simplicity, we assume here that ${\boldsymbol D}^{(a)}_{12} = {\boldsymbol D}^{(b)}_{12} = {\boldsymbol D}_{12}$ and that ${\boldsymbol D}_{12}$ is a real vector, while $\hat {\boldsymbol x}$ is a unit vector pointing in the direction of the positive $x$ axis. When $|\hat {\boldsymbol D}_{12} \cdot \hat {\boldsymbol x}| = 1$, both atomic dipole moments are orthogonal to the mirror surface in the $x=0$ plane. As one would expect, the mirror mediated interactions between the atoms are relatively weak in this case. These assume their maximum, when ${\boldsymbol D}_{12}$ is parallel to the mirror surface and $|\hat {\boldsymbol D}_{12} \cdot \hat {\boldsymbol x}| = 0$. (b) The real part of $\Gamma_{\rm mir}^{(ab)}$ in Eq.~(\ref{B17}) which represents changes to the spontaneous decay rates as a function of $\xi$ and for different values of $|\hat {\boldsymbol D}_{12} \cdot \hat {\boldsymbol x}|$.}
	\label{fig3}
\end{figure*}

\subsection{Dicke sub- and superradiance}

Taking into account that some of the light that has been emitted by each atom travels to the opposite side of the mirror interface where it interferes with the reflected light originating from the atom on the opposite side, one can show that the operators $\mathcal{L} (\rho_{\rm A})$ and $H_{\rm cond}$ in Eq.~(\ref{eq:rho30}) can be written as
\begin{eqnarray} \label{L}
\mathcal{L} (\rho_{\rm A}) &=& \sum_{i,j = a,b} {\rm Re} \left( \Gamma_{\rm mir}^{(ij)} \right) \, \sigma_{i}^{-} \rho_{\rm A} \sigma_{j}^{+} \, , \nonumber \\ 
H_{\rm cond} &=& H_{\rm A} - \frac{{\rm i} \hbar}{2} \sum_{i,j=a,b} \Gamma_{\rm mir}^{(ij)} \, \sigma_{j}^{+} \sigma_{i}^{-} \, .
\end{eqnarray}
The constants $\Gamma_{\rm mir}^{(ij)}$ in these equations depend on the properties of the atoms and on the average reflection and transmission rates $t_i$ and $r_i$ of the mirror interface. In the following, we assume that these do not depend on the angle of incidence and the frequency of the incoming light. Such a dependence would alter the strength of the predicted interactions but we expect that our results remain valid also in the more general case, at least qualitatively.

Here we are especially interested in the case where the distance between atom $a$ the mirror image of atom $b$, i.e.~the difference between  $\boldsymbol{r}_a = (x_a,y_a,z_a)$ and $\tilde{\boldsymbol{r}}_b = (-x_b,y_b,z_b)$, is relatively small. For simplicity, let us assume that $y_a = y_b$ and $z_a = z_b$ such that the relative effective distance $\xi = k_0 \| \boldsymbol{r}_a - \tilde{\boldsymbol{r}}_b \|$ equals $k_0 (x_a + x_b)$. Using this notation and considering real mirror transmission and reflection rates for simplicity, one can show that
\begin{eqnarray} \label{B17}
\Gamma_{\rm mir}^{(ab)} &=& {3 \over 8} t_a r_b  \Gamma_{\rm free} \left[  \hat {\boldsymbol D}^{(a)}_{12} \cdot \hat {\boldsymbol D}^{(b)}_{12} \left( {1 \over {\rm i} \xi} + {1 \over \xi^2} - {1 \over {\rm i} \xi^3} \right) \right. \notag \\
&& - \left. \left( \hat {\boldsymbol D}^{(a)}_{12} \cdot \hat {\boldsymbol x} \right) \left( \hat {\boldsymbol D}^{(b)}_{12} \cdot \hat {\boldsymbol x} \right) \left( {1 \over {\rm i} \xi} + {3 \over \xi^2} - {3 \over {\rm i} \xi^3} \right) \right] {\rm e}^{{\rm i} \xi} \notag \\
&& - {3 \over 16} t_a r_b  \Gamma_{\rm free} \left[ \hat {\boldsymbol D}^{(a)}_{12} \cdot \hat {\boldsymbol D}^{(b)}_{12} \left( {1 \over {\rm i} \xi} - {2 \over {\rm i} \xi^3} \right) \right. \notag \\
&& \left. + \left( \hat {\boldsymbol D}^{(a)}_{12} \cdot \hat {\boldsymbol x} \right) \left( \hat {\boldsymbol D}^{(b)}_{12} \cdot \hat {\boldsymbol x} \right) \left( {1 \over {\rm i} \xi} + {6 \over {\rm i} \xi^3} \right) \right] \, , 
\end{eqnarray}
while $\Gamma^{(aa)}_{\rm mir} = \Gamma^{(bb)}_{\rm mir} = \Gamma_{\rm free}$ and $\Gamma_{\rm mir}^{(ba)}  = \Gamma_{\rm mir}^{(ab)*}$. Here $\Gamma_{\rm free}$ denotes the single-atom free space decay rate and $\hat{\boldsymbol D}^{(i)}_{12} $ and $\hat{\boldsymbol x}$ are unit vectors which point in the direction of the (real) dipole moment vector $\boldsymbol D^{(i)}_{12}$ and of the positive $x$ axis, respectively. 

As we shall see below, the real part ${\rm Re}(\Gamma_{\rm mir}^{(ab)})$ of the complex rate in Eq.~(\ref{B17}) results in corrections to the spontaneous decay rate of certain symmetric atomic state, while its imaginary part  $\Delta_{\rm mir} = {\rm Im}(\Gamma_{\rm mir}^{(ab)})$ describes level shifts. Fig.~\ref{fig3} shows both frequencies for different orientations of the atomic dipole moments and for different effective relative distances $\xi$ between atom $a$ and the mirror image of atom $b$. The rate $\Gamma_{\rm mir}^{(ab)}$ is only non-zero when $\xi$ is comparable to the wavelength of the emitted light, however, the actual distance of the atoms can be much larger. In the absence of a mirror interface, the reflection rate $r_b =0$ and $\Gamma_{\rm mir}^{(ab)}$ tends to zero, as one would expect. Formally, Eq.~(\ref{eq:rho30}) is exactly the same as the master equations of two atoms experiencing Dicke sub- and superradiance \cite{Agarwal,Haroche,new1,Ficek}. The only difference is the overall factor $ {3 \over 8} t_a r_b $ in Eq.~(\ref{B17}). In addition, there are some additional imaginary terms in the third and fourth line of this equation.

\section{Discussion}

\subsection{Alterations of atomic level shifts and spontaneous decay rates}

Having a closer look at the conditional Hamiltonian $H_{\rm cond}$ in Eq.~(\ref{L}), we see that it contains a Hermitian and a non-Hermitian contribution. The Hermitian contribution contains the atom Hamiltonian $H_{\rm A}$ and terms proportional to $\Delta_{\rm mir} = {\rm Im}(\Gamma_{\rm mir}^{(ab)})$. These describe the free dynamics of the atoms as well as interaction-induced level shifts. As one can see when comparing Eq.~(\ref{B17}) with the equations in Refs.~\cite{Dicke,Agarwal,new1,Haroche,Ficek}, the level shifts in the first two lines of Eq.~(\ref{B17}) are essentially the same as the level shifts in the presence of free-space dipole-dipole interactions between two two-level atoms at positions ${\boldsymbol r}_a$ and $\tilde {\boldsymbol r}_b$. The only difference is the above mentioned overall factor, which occurs since not all emitted light contributes to the generation of the interaction. In addition there are some additional level shifts in the last two lines of Eq.~(\ref{B17}). However, as illustrated in Fig.~\ref{fig3}(a), these do not significantly alter the general dependence of the level shifts on the relative effective distance $\xi$ of the two atoms. For example, when $\xi$ tends to zero, $\Delta_{\rm mir}$ tends to infinity due to the above quantum optical model treating the atoms as point particles. Because of these similarities, we refer to the mirror-mediated interaction which we predict here as a dipole-dipole interaction.

The remaining terms in the conditional Hamiltonian $H_{\rm cond}$ in Eq.~(\ref{L}) describe the damping of population in excited atomic states. By diagonalising $H_{\rm cond}$ we find that the spontaneous decay rate of the state $|22 \rangle$ of the two two-level atoms with both atoms in their excited state equals $2 \Gamma_{\rm free}$, as usual. However, collective atomic states which share only one excitation now have the spontaneous decay rates 
\begin{eqnarray} \label{rates}
\Gamma_\pm = \Gamma_{\rm free} \pm {\rm Re} (\Gamma^{(ab)}_{\rm mir}) \, .
\end{eqnarray}
As we can see from Eq.~(\ref{B17}) and Fig.~\ref{fig3}(b), up to an overall factor, the differences between $\Gamma_\pm$ and $\Gamma_{\rm free}$ are what  they would be in the presence of a free-space dipole-dipole interaction between two atoms at positions ${\boldsymbol r}_a$ and $\tilde {\boldsymbol r}_b$ \cite{Dicke,Agarwal,Haroche,new1,Ficek}. As shown in Section \ref{appD}, the atomic states with well-defined spontaneous decay rates are the same as for dipole-interacting atoms, namely the double-excited state $|22 \rangle$ and the single-excited symmetric and antisymmetric states $|\pm \rangle = (|12 \rangle \pm |21 \rangle)/\sqrt{2}$.

Changes to spontaneous decay rates can be detected, for example, with the help of fluorescence lifetime measurements. Moreover, when the atoms are driven by a common laser field, we expect their higher order photon correlation functions \cite{BeHe,Antoine,Schmidt-Kaler} to change and an interference pattern to emerge, if the spontaneously emitted photons are collected on a far-away screen, as illustrated in Fig.~\ref{fig1}(b). As described in Methods, the only assumptions regarding the distance of the two two-level atoms made in the derivation of the above equations are: 
\begin{enumerate}
\item The actual distance between the atoms and between an atom and the mirror interface should be relatively large. This applies when $k_0 |x_a - x_b| \gg 1$. 
\item The actual distance between the two atoms in Figs.~\ref{fig1}(b) and (c) should not be so large that the time it takes light to travel from one atom to the other becomes comparable to the lifetime of excited atomic states.
\end{enumerate}
The first condition allows us to ignore direct atom-atom and atom-mirror interactions which are relatively short-range. The second condition simplifies the modelling of light propagation in the presence of the mirror interface and is not very restrictive. For example, light can travel a $1 \,$mm distance in less than $3.4 \cdot 10^{-12} \,$s which is much shorter than the typical lifetime $1/\Gamma_{\rm free}$ of excited atomic states. However, when analysing atomic interactions over very large distances, retardation effects need to be taken into account and the dynamics of the two atoms can no longer be described by a simple Markovian master equation, like the one in Eq.~(\ref{eq:rho30}). 

\begin{figure*}[t] 
	\centering
	\includegraphics[width=0.98 \textwidth]{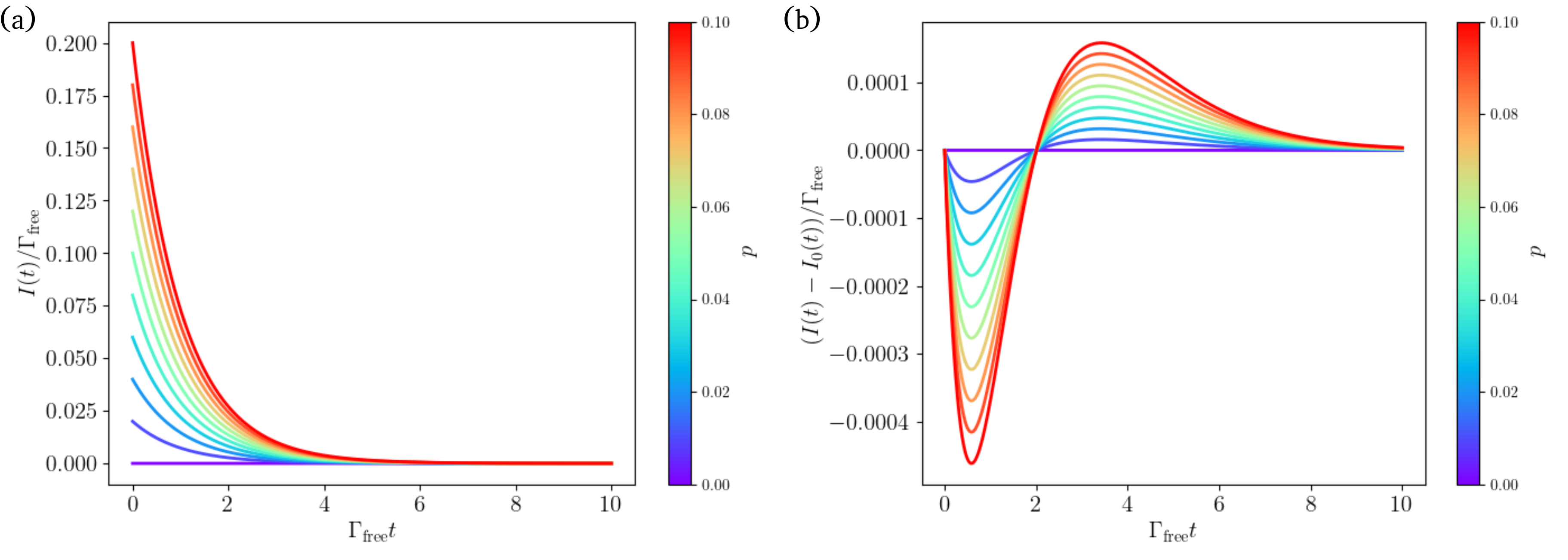} 
\caption{(a) Time dependence of the photon emission rate $I(t)$ in Eq.~(\ref{C5}) for different initial populations $p$ of the excited atomic state in the presence of mirror-mediated dipole-dipole interactions. Here ${\rm Re} \big( \Gamma_{\rm mir}^{(ab)} \big) = 0.05 \, \Gamma_{\rm free}$. (b) The rate $I(t)$ differs from the emission rate $I_0(t)$ of the atoms in the absence of interactions for the same $p$. As one would expect in the case of a broadening of spontaneous decay rates, the loss of atomic excitation happens faster at relatively short times and slower at later times.}
	\label{fig4}
\end{figure*}

The main difference between the above-described mirror-mediated dipole-dipole interactions and the usual dipole-dipole interactions between two atoms in free space is that the former can be felt over much longer distances. As we have shown above, mirror-mediated remote dipole-dipole interaction can persist over distances which are several orders of magnitude longer than the wavelength of the emitted light. They assume a maximum when the relative distance $\xi$ between the position $\boldsymbol{r}_a$ of atom $a$ and the position $\tilde{\boldsymbol{r}}_b$ of the mirror image of atom $b$ is of the order of one. This can be the case even when the actual distance $\| \boldsymbol{r}_a - \boldsymbol{r}_b \|$ of the atoms is several orders of magnitude larger than the wavelength of the emitted light. The interaction which we predict here is therefore ultralong-range and targeted. 

Another requirement for the atomic interactions which we predict in this paper is the presence of an {\em asymmetric} mirror interface. As illustrated in Fig.~\ref{fig1}, such a mirror can be realised with the help of a different surface roughness on both sides of the reflecting layer. If both sides of the mirror surface were smooth, the interaction constant $\Gamma_{\rm mir}^{(ab)} $ in Eq.~(\ref{B17}) would be proportional to $r_a^* t_b +  t_a^* r_b $ which is zero, as we know from classical optics \cite{Abeer}. For mirrors which are equally smooth on both sides, the interaction which we predict here therefore simply disappears. However, for the asymmetric mirror interface shown in Figs.~\ref{fig1}(b) and (c), the predicted atomic interaction scales as $t_a^* r_b$ which is in general non-zero.

\subsection{Predictions for fluorescence lifetime measurements} \label{appD}

To determine the spontaneous decay rates of two atoms with ground states $|1 \rangle_i$ and excited states $|2\rangle_i$ on opposite sides of a partially transparent mirror interface, we absorb all the Hermitian terms of the conditional Hamiltonian $H_{\rm cond}$ in Eq.~(\ref{L}) into the free atomic Hamiltonian $H_{\rm A}$. Doing so, $H_{\rm cond}$ can be written as
\begin{eqnarray} \label{M}
H_{\rm cond} = H_{\rm A} - {{\rm i} \hbar \over 2} \left[ \Gamma_+ \, L_+^\dagger L_+ + \Gamma_- \, L_-^\dagger L_- \right] 
\end{eqnarray}
where the $\Gamma_\pm$ are the spontaneous decay rates of the two atoms in Eq.~(\ref{rates}) and where the $L_\pm$ with
\begin{eqnarray}
 L_\pm &=& (\sigma_a^- \pm \sigma_b^-)/\sqrt{2} 
 \end{eqnarray} 
 are atomic lowering operators. Hence the time evolution operator $U_{\rm cond}(t,0) = \exp \left( - {\rm i} H_{\rm cond} t /\hbar \right) $ which describes the dynamics of atom $a$ and atom $b$ under the condition of no photon emission in $(0,t)$ equals, in the interaction picture with respect to $H_0 = H_{\rm A}$ and $t=0$,
 \begin{eqnarray} \label{Ucond}
U_{\rm cond}(t,0) &=& |+ \rangle \langle +| \, {\rm e}^{- \Gamma_+ t/2} +  |- \rangle \langle -| \, {\rm e}^{- \Gamma_- t/2}  \notag \\
&& + |11 \rangle \langle 11| + |22 \rangle \langle 22| \, {\rm e}^{- \Gamma_{\rm free} t} \, .
\end{eqnarray}
with $|\pm \rangle = (|12\rangle \pm |21\rangle /\sqrt{2}$, as mentioned already in the beginning of this section.

Suppose an incoherent excitation process prepares each atom with probability $p$ in its excited state, thereby creating a statistical mixture of the atomic states $|11 \rangle$, $|12 \rangle$, $|21 \rangle$ and $|22 \rangle$. In this case, the probability $P_0(t) = \| U_{\rm cond}(t,0) \, |\psi_{\rm I} \rangle \|^2$ for no photon emission in $(0,t)$ is the sum of three exponentials and equals \cite{Hegerfeldt,Stokes}
\begin{eqnarray} \label{I(t)}
P_0(t) &=& (1-p)^2 + (1-p) p \left( {\rm e}^{- \Gamma_+ t} + {\rm e}^{- \Gamma_- t} \right) \notag \\
&& + p^2 \, {\rm e}^{- 2 \Gamma_{\rm free}t} \, . 
\end{eqnarray}
For $p \ll 1$, the probability of finding both atoms in the excited state becomes negligible and the probability density $I(t)$ for a photon emission at $t$ coincides with the probability density $w_1(t) = - {\rm d}/ {\rm d}t \, P_0(t)$ for the emission of a first photon at $t$. Hence, 
\begin{eqnarray} \label{C5}
I(t) &=& 2p \left[ \Gamma_{\rm free} \, {\rm cosh} \left( {\rm Re} \big( \Gamma_{\rm mir}^{(ab)} \big) t \right) \right. \notag \\
&& \left. - {\rm Re} \big( \Gamma_{\rm mir}^{(ab)} \big) \, {\rm sinh} \left( {\rm Re} \big( \Gamma_{\rm mir}^{(ab)} \big) t \right) \right]  {\rm e}^{ - \Gamma_{\rm free} t} 
\end{eqnarray}
to a very good approximation. This equation holds up to first order in $p$. As illustrated in Fig.~\ref{fig4}, this emission rate is qualitatively different from the emission rate $I_0(t)$ of the atoms in the absence of dipole-dipole interactions. It is therefore possible to use fluorescence lifetime measurements to detect the above described changes of spontaneous decay rates and to obtain a signature of the remote mirror-mediated dipole-dipole interactions which we predict in this paper.

\section{Conclusions}

In this paper we derived the quantum optical master equations of two two-level atoms on opposite sides of a partially transparent asymmetric mirror interface by evolving the atoms and the free radiation field for a short time interval $\Delta t$ using second order perturbation theory. Our approach allows us to deduce the time derivative of the atomic density matrix ${\rho}_{\rm A}$ from the {\em classical} dynamics of light in the {\em absence} of any atom-field interactions. We then showed that the two atoms can experience an effective dipole-dipole interaction when atom $a$ is close to the position of the mirror image of atom $b$ and vice versa. The main result of this paper is the prediction of targeted, remote, mirror-mediated ultralong-range dipole-dipole interactions which are likely to find a wide range of applications in the design of novel photonic devices for quantum technology applications, like non-invasive quantum sensing with fluorescence lifetime measurements.

\section{Methods}

Our starting point for the derivation of the quantum optical master equations in Eq.~(\ref{eq:rho30}) is the system Hamiltonian $H$ which can be written as 
\begin{eqnarray} \label{H}
H &=& H_{\rm A} + H_{\rm F} + H_{\rm AF} \, .
 \end{eqnarray}
Here $H_{\rm A}$ and $H_{\rm F}$ denote the free energy of the atoms and of the electromagnetic field, i.e.~in the absence of the mirror interface. An expression for the interaction Hamiltonian $H_{\rm AF}$ between the atoms and the local excitations of the surrounding free radiation field within the usual dipole approximation can be found in Eq.~(\ref{eq:Hdef2}) \cite{Agarwal,Haroche,new1,Ficek}. As we shall see below, in addition, we only need to know how the atomic dipole moments and electric field observables evolve in the Heisenberg picture in the {\em absence} of atom-field interactions.

Suppose $\rho_{\rm A}(0)$ is the initial density matrix of the two atoms in the Schr\"odinger picture, while the surrounding free radiation field is initially in its vacuum state. We then evolve the atom-field density matrix $|0_{\rm F} \rangle \rho_{\rm A}(0) \langle 0_{\rm F}|$ for a time $\Delta t$ with the time evolution operator $U(\Delta t,0)$ of the above Hamiltonian $H$. Subsequently performing an absorbing measurement on the surrounding free radiation field leads to the atomic density matrix
\begin{eqnarray} \label{BB1}
\rho_{\rm A}(\Delta t) = {\rm Tr}_{\rm F} \big[ U(\Delta t,0) |0_{\rm F} \rangle \rho_{\rm A}(0) \langle 0_{\rm F}| U^\dagger(\Delta t,0) \big] \, .
\end{eqnarray}
Here the trace over the field degrees of freedom is taken to ensure that a measurement on the surrounding electromagnetic field does not change the properties of the atoms, if its outcome is ignored. As requested by locality, the density matrices $\rho_{\rm A}(\Delta t) $ and $U(\Delta t,0) |0_{\rm F} \rangle \rho_{\rm A}(0) \langle 0_{\rm F}| U^\dagger(\Delta t,0)$ must have the same atomic expectation values. Next, we introduce the time derivative 
\begin{eqnarray} \label{BB2}
\dot {\rho}_{\rm A} = {1 \over \Delta t} \left( \rho_{\rm A}(\Delta t) - \rho_{\rm A}(0) \right)  
\end{eqnarray}
which describes the dynamics of the atomic density ${\rho}_{\rm A}$ on the coarse grained time scale $\Delta t$, while the free radiation field at the position of the atoms remains effectively in its vacuum state \cite{Hegerfeldt,Stokes}.

Since the time evolution operator $U(\Delta t,0)$ in Eq.~(\ref{BB1}) cannot be calculated easily analytically, we write the total Hamiltonian $H$ of the experimental setup in Fig.~\ref{fig1}(c) in the following as the sum of the free Hamiltonian $H_0 = H_{\rm A} + H_{\rm F}$ and the interaction $H_{\rm AF}$.  As long as $\Delta t$ is neither too long nor too short, as described in Results, we can analyse the dynamics of the system using a Dyson series expansion which implies that
\begin{widetext}
\begin{eqnarray} \label{BB3}
U(\Delta t,0) &=& U_0(\Delta t,0) - {{\rm i} \over \hbar} \int_0^{\Delta t} {\rm d}t \, U_0(\Delta t,t) \, H_{\rm AF} \, U_0(t,0) 
- {1 \over \hbar^2} \int_0^{\Delta t} {\rm d}t \int_0^t {\rm d}t' \, U_0(\Delta t,t) \, H_{\rm AF} \, U_0(t,t') \, H_{\rm AF} \, U_0(t',0) \nonumber \\
\end{eqnarray}
to a very good approximation. Combining Eqs.~(\ref{BB1}) and (\ref{BB3}), while only taking terms in zeroth order in $\Delta t$ into account, leads to 
\begin{eqnarray} \label{BB40}
\rho_{\rm A}(\Delta t) &=& \frac{1}{\hbar^{2}} \int_{0}^{\Delta t} {\rm d}t \int_0^{\Delta t} {\rm d}t' \, {\rm Tr}_{\rm F} \left[ U_0(\Delta t,t) \, H_{\rm AF} \, U_0(t,0) \, |0_{\rm F} \rangle \rho_{\rm A}(0) \langle 0_{\rm F}| \, U^\dagger_0(t',0) \, H_{\rm AF} \, U^\dagger_0(\Delta t,t') \right] \nonumber \\
&& - \frac{1}{\hbar^2} \int \limits_{0}^{\Delta t} {\rm d}t \int \limits_0^{t} {\rm d}t' \, \langle 0_{\rm F}|  U_0(\Delta t,t) \, H_{\rm AF} \, U_0(t,t') \, H_{\rm AF} \, U_0(t',0) |0_{\rm F} \rangle \rho_{\rm A}(0) + {\rm c.c.} \nonumber \\
&&  + \langle 0_{\rm F}| U_0 (\Delta t,0) |0_{\rm F} \rangle \rho_{\rm A}(0) \langle 0_{\rm F}| U^\dagger_0 (\Delta t,0) |0_{\rm F} \rangle   
\end{eqnarray}
which applies in first order in $\Delta t$. To obtain the above equation, we took into account that $H_{\rm AF}$ either creates or annihilates a photon, while $H_0$ preserves the number of excitations in the free radiation field. Carefully comparing this equation with Eqs.~(\ref{eq:rho30}) and (\ref{BB2}), we see that 
\begin{eqnarray} \label{BB4}
{\cal L} (\rho_{\rm A}) &=& \frac{1}{\hbar^{2} \Delta t}  \int_{0}^{\Delta t} {\rm d}t \int_0^{\Delta t} {\rm d}t' \, {\rm Tr}_{\rm F} \left[ U_0(\Delta t,t) \, H_{\rm AF} \, U_0(t,0) \, |0_{\rm F} \rangle \rho_{\rm A} \langle 0_{\rm F}| \, U^\dagger_0(t',0) \, H_{\rm AF} \, U^\dagger_0(\Delta t,t') \right] \, , \nonumber \\
H_{\rm cond} &=& H_{\rm A} - \frac{{\rm i}}{\hbar \Delta t} \int \limits_{0}^{\Delta t} {\rm d}t \int \limits_0^{t} {\rm d}t' \, \langle 0_{\rm F}|  U_0(\Delta t,t) \, H_{\rm AF} \, U_0(t,t') \, H_{\rm AF} \, U_0(t',0) |0_{\rm F} \rangle \, .
\end{eqnarray}
To further simplify the above expressions, we notice that $H_0$ is the sum of two commuting Hamiltonians, namely $H_{\rm A}$ and $H_{\rm F}$. Hence, $U_0(t,0) = U_{\rm A}(t,0) \otimes U_{\rm F} (t,0)$, where $U_{\rm A} (t,0)$ and $U_{\rm F} (t,0)$ denote the time evolution operators associated with $H_{\rm A}$ and $H_{\rm F}$, respectively. In addition, we introduce the short hand notation  
\begin{eqnarray} \label{BB11}
\boldsymbol{D}^{(i)}(t) = U^\dagger_{\rm A}(t,0) \, \boldsymbol{D}_{12}^{(i)} \sigma_i^- \, U_{\rm A}(t,0) 
\end{eqnarray}
and notice that the vacuum state is invariant under $U_{\rm F}$. Hence, using Eqs.~(\ref{eq:Hdef2}) and (\ref{BB4}), one can show that 
\begin{eqnarray} \label{BB5}
{\cal L} (\rho_{\rm A}) &=& \frac{e^2}{\hbar^{2} \Delta t} \, \sum_{i,j=a,b} \int_{0}^{\Delta t} {\rm d}t \int_0^{\Delta t} {\rm d}t' \, {\rm Tr}_{\rm F} \left[ \boldsymbol{D}^{(i)}(t) \cdot U_{\rm F}(\Delta t,t) \boldsymbol{\cal E}(\boldsymbol{r}_i) \, |0_{\rm F} \rangle \rho_{\rm A} \langle 0_{\rm F}| \, \boldsymbol{D}^{(j)}(t')^\dagger \cdot \boldsymbol{\cal E}(\boldsymbol{r}_j)^\dagger \, U^\dagger_{\rm F}(\Delta t ,t') \right] , \nonumber \\
H_{\rm cond} &=& H_{\rm A} - \frac{{\rm i} e^2}{\hbar \Delta t} \, \sum_{i,j=a,b} \int \limits_{0}^{\Delta t} {\rm d}t \int \limits_0^{t} {\rm d}t' \, 
\boldsymbol{D}^{(j)}(t)^\dagger \cdot \langle 0_{\rm F}| \boldsymbol{\cal E} (\boldsymbol{r}_j)^\dagger \, U^\dagger_{\rm F}(t',0) \, 
\boldsymbol{D}^{(i)}(t') \cdot U_{\rm F}(t,0) \boldsymbol{\cal E}(\boldsymbol{r}_i) \, |0_{\rm F} \rangle 
\end{eqnarray}
in zeroth order in $\Delta t$. Here ${\cal L} (\rho_{\rm A}) $ contains all the contributions of the atom-field density matrix which correspond to the presence of a photon at $\Delta t$ in the free radiation field. It therefore equals the (unnormalised) density matrix of the atoms conditional on the creation of a photon in $(0,\Delta t)$. Analogously, the non-Hermitian Hamiltonian $H_{\rm cond}$ only contains contributions in which excitation has been created in $(0,\Delta t)$ but is later re-absorbed by the atoms. Hence it describes atomic dynamics in the absence of an emission \cite{Hegerfeldt,Stokes}.

\subsection{The free-space dynamics of atoms and field} \label{appA}  

Suppose $\hbar \omega_0$ is the energy gap between the ground and the excited state of atom $i$. Then the atom Hamiltonian $H_{\rm A}$ can be written as $H_{\rm A} = \sum_{i=a,b} \hbar \omega_0 \, \sigma^+_{i} \sigma^-_{i}$ with $\sigma^{+}_{i} = |2 \rangle_{ii} \langle 1|$ and $\sigma^{-}_{i} = |1 \rangle_{ii} \langle 2|$. Hence the time-dependent dipole moment operator $ \boldsymbol{D}^{(i)}(t)$ in Eq.~(\ref{BB11}) equals
\begin{eqnarray} \label{eq:Hdef3}
\boldsymbol{D}^{(i)}(t) = {\rm e}^{- {\rm i} \omega_0 t} \, \boldsymbol{D}_{12}^{(i)} \, \sigma^{-}_{i} \, .
\end{eqnarray}
From Eq.~(\ref{BB5}) we see that the only other expression needed for the derivation of the quantum optical master equations in Eq.~(\ref{eq:rho30}) is the state $U_{\rm F}(t,0) \, \boldsymbol{\cal E}_{\boldsymbol{s} \lambda}(\boldsymbol{r}) \, |0_{\rm F} \rangle$ of the free radiation field. This state is obtained when creating a local field excitation with direction of propagation $\boldsymbol{s}$ and polarisation $\lambda$ at time $t=0$ at position $\boldsymbol{r}$ and subsequently evolving the resulting state for some time $t$. Since $\Delta t$ is much larger than the time it takes light to travel from the atoms to the mirror surface, light emitted at $t=0$ in the direction of the mirror has either already been reflected or transmitted after almost all times $t \in (0,\Delta t)$. Neglecting very small times $t$ for which light has not yet reached the mirror interface and using Eqs.~(\ref{eq:dyn}) and (\ref{Efieldinterfacenew2}), we therefore find that
\begin{eqnarray} \label{eq:rho3zzzz}
U_{\rm F}(t,0) \, \boldsymbol{\cal E}_{\boldsymbol{s} \lambda} (\boldsymbol{r}_a) \, |0_{\rm F} \rangle
&=& \Theta(-s_x) \left[ r_a(\boldsymbol{s}) \, \boldsymbol{\cal E}_{{\boldsymbol S}({\boldsymbol s}) \lambda} ({\boldsymbol R}_a({\boldsymbol s},t)) + t_a(\boldsymbol{s}) \, \boldsymbol{\cal E}_{\boldsymbol{s} \lambda} (\boldsymbol{r}_a + \boldsymbol{s} c t) \right] |0_{\rm F} \rangle 
+ \Theta(s_x) \, \boldsymbol{\cal E}_{\boldsymbol{s} \lambda} (\boldsymbol{r}_a + \boldsymbol{s} c t) \, |0_{\rm F} \rangle \, , \nonumber \\
U_{\rm F}(t,0) \, \boldsymbol{\cal E}_{\boldsymbol{s} \lambda}(\boldsymbol{r}_b) \, |0_{\rm F} \rangle
&=& \Theta(s_x) \left[ r_b(\boldsymbol{s}) \, \boldsymbol{\cal E}_{\tilde{\boldsymbol{s}} \lambda} (\tilde{\boldsymbol{r}}_b + \tilde{\boldsymbol{s}} c t)
+ t_b(\boldsymbol{s}) \, \boldsymbol{\cal E}_{\boldsymbol{s} \lambda} (\boldsymbol{r}_b + \boldsymbol{s} c t) \right] |0_{\rm F} \rangle 
+ \Theta(-s_x) \, \boldsymbol{\cal E}_{\boldsymbol{s} \lambda} (\boldsymbol{r}_b + \boldsymbol{s} c t) \, |0_{\rm F} \rangle 
\end{eqnarray}
for direction vectors $\boldsymbol{s} = (s_x,s_y,s_z)$. Here the Heavyside function $\Theta(s_x)$ equals 0 for $s_x < 0$ and 1 otherwise. Moreover, the real reflection rates $r_i(\boldsymbol{s})$ for light travelling from atom $i$ in the $\boldsymbol{s}$ direction either equal $0$ or $1$, depending on whether light arrives at a metallic island or at a gap in the mirror interface (cf.~Fig.~\ref{fig1}). The corresponding transmission rates $t_i(\boldsymbol{s})$ are given by $t_i(\boldsymbol{s}) = 1 - r_i(\boldsymbol{s})$, since light with a well defined direction of propagation and source cannot be reflected {\em and} transmitted by the mirror surface. Later on, we will take into account that the effective reflection and transmission rates of the mirror are given by 
\begin{eqnarray} \label{oma100}
r_i = {1 \over 2\pi} \int_{{\cal S}_i} {\rm d} {\boldsymbol s} \, r_i(\boldsymbol{s})  ~~~ {\rm and}  ~~~ t_i = 1-r_i \, ,
\end{eqnarray}
where ${\cal S}_a = \{ {\boldsymbol s} \in {\cal S}: s_x<0 \}$ and ${\cal S}_b = \{ {\boldsymbol s} \in {\cal S}: s_x>0 \}$. As previously mentioned in the Results section, ${\boldsymbol S}({\boldsymbol s})$ and ${\boldsymbol R}_a({\boldsymbol s},t)$ denote the direction of propagation and the position of a local electric field excitation at time $t$ after its creation by atom $a$ at $t=0$ and after its subsequent reflection on the rough side of the mirror interface.

\subsection{The conditional Hamiltonian $H_{\rm cond} $}

Substituting Eq.~(\ref{eq:Hdef3}) into Eq.~(\ref{BB5}), we can show that the conditional Hamiltonian $H_{\rm cond}$ can indeed be written as in Eq.~(\ref{L}), if we define the constants $\Gamma_{\rm mir}^{(ij)}$ such that
\begin{eqnarray} \label{B8}
\Gamma_{\rm mir}^{(ij)} = {1 \over \Delta t} \int_0^{\Delta t} {\rm d} t \int_0^t {\rm d} t' \, {e^2c \over 8\hbar  \pi^3 \varepsilon} \, {\rm e}^{{\rm i} \omega_0 (t - t')} \, \gamma_{\rm mir}^{(ij)} (t,t') 
\end{eqnarray}
with $\gamma_{\rm mir}^{(ij)} (t,t')$ given by
\begin{eqnarray} \label{B9} 
\gamma_{\rm mir}^{(ij)} (t,t') = {16 \pi^3 \varepsilon \over \hbar c} \, \boldsymbol{D}_{12}^{(j)} \cdot \langle 0_{\rm F}| \boldsymbol{\cal E} (\boldsymbol{r}_j)^\dagger \, U^\dagger_{\rm F}(t',0) \, \boldsymbol{D}^{(i)}_{12} \cdot U_{\rm F}(t,0) \boldsymbol{\cal E}(\boldsymbol{r}_i) |0_{\rm F} \rangle \, .
\end{eqnarray}
Using Eqs.~(\ref{eq:observables}), (\ref{eq:rho3xxxxx}) and (\ref{eq:rho3zzzz}) and performing one of the ${\boldsymbol k}$ integrations, we therefore find that
\begin{eqnarray} \label{B10}
\gamma_{\rm mir}^{(aa)} (t,t')
&=& \sum_{\lambda=1,2} \int {\rm d} \boldsymbol{k} \, k \, {\rm e}^{- {\rm i} ck (t-t')}  
\left[ \left( t_a({\boldsymbol s}) \, \Theta (-s_x) + \Theta (s_x) \right) \big( \boldsymbol{D}^{(a)}_{12} \cdot  \boldsymbol{e}_{\boldsymbol{s} \lambda} \big)^2 
+ r_a({\boldsymbol s}) \, \Theta(-s_x) \, \big( \boldsymbol{D}^{(a)}_{12} \cdot \boldsymbol{e}_{S({\boldsymbol s}) \lambda} \big)^2 \right]  \notag \\
\gamma_{\rm mir}^{(bb)} (t,t')
&=& \sum_{\lambda=1,2} \int {\rm d} \boldsymbol{k} \, k \, {\rm e}^{- {\rm i} ck (t-t')}  
\left[ \left( \Theta (-s_x) + t_b({\boldsymbol s}) \, \Theta (s_x) \right) \big( \boldsymbol{D}^{(b)}_{12} \cdot \boldsymbol{e}_{\boldsymbol{s} \lambda} \big)^2  
+  r_b({\boldsymbol s}) \, \Theta (s_x) \, \big( \boldsymbol{D}^{(b)}_{12} \cdot \boldsymbol{e}_{\tilde{\boldsymbol{s}} \lambda} \big)^2 \right] 
\end{eqnarray}
with ${\rm d}\boldsymbol{k} = {\rm d} \boldsymbol{s} \, {\rm d}k \, k^2$. Next we introduce polar coordinates $k \in (0,\infty)$, $\varphi \in (0,2\pi)$ and $\vartheta \in (0,\pi)$ such that
\begin{eqnarray} \label{vectors}
\boldsymbol{s} = \begin{pmatrix} \cos \vartheta \\ \cos \varphi \, \sin \vartheta \\ \sin \varphi \, \sin \vartheta \end{pmatrix} , ~~~
\boldsymbol{e}_{\boldsymbol{s} 1} = \begin{pmatrix} 0 \\ \sin\varphi \\ - \cos\varphi \end{pmatrix} , ~~~ 
\boldsymbol{e}_{\boldsymbol{s} 2} = \begin{pmatrix} \sin\vartheta\\ - \cos\varphi\cos\vartheta\\ - \sin\varphi\cos\vartheta \end{pmatrix} , 
\end{eqnarray}
while ${\rm d}\boldsymbol{s} = {\rm d} \vartheta \, {\rm d} \varphi  \, \sin \vartheta$. After replacing the reflections and transmission rates $r_i(\boldsymbol{s})$ and $t_i(\boldsymbol{s})$ by their average values $r_i$ and $t_i$ in Eq.~(\ref{oma100}), which is well justified when the metallic islands which form the mirror interface are much smaller then the wavelength of the emitted light, Eq.~(\ref{B10}) contains the integral
\begin{eqnarray}\label{B11}
\int_0^\infty {\rm d}k \, k^3 \, {\rm e}^{-{\rm i} c k \tau} = - {{\rm i} \pi  \over c^4} \, \delta^{(3)} (\tau) 
\end{eqnarray}
with $\delta^{(3)} (\tau) $ denoting the third derivative of $\delta(\tau)$ with respect to $\tau$. Hence we can now show that 
\begin{eqnarray} \label{B13}
{1 \over \Delta t} \int_0^{\Delta t} {\rm d} t \int_0^t {\rm d} t' \, {\rm e}^{{\rm i} \omega_0 (t - t')} \int_0^\infty {\rm d} k \, k^3 \, {\rm e}^{- {\rm i} ck (t-t')}
= {{\rm i} \pi \over c^3 \Delta t} \int_0^{\Delta t} {\rm d}t \int_0^t {\rm d} \tau \, {\rm e}^{{\rm i} \omega_0 \tau} \, \delta^{(3)} (\tau)
= {\pi \omega_0^3 \over c^4} \, .
\end{eqnarray}
Combining the above equations and assuming that the direction vectors ${\boldsymbol S}({\boldsymbol s})$ cover the half-space on the right hand side of the mirror interface evenly, we then find that 
\begin{eqnarray} \label{Bii}
\Gamma_{\rm mir}^{(ii)} = {e^2 \omega_0^3 \over 8  \pi^2 \hbar \varepsilon c^3} \sum_{\lambda=1,2} \int {\rm d} \boldsymbol{s} \, 
\left[ r_i \, \Theta(\mp s_x) \, \big( \boldsymbol{D}^{(i)}_{12} \cdot \boldsymbol{e}_{\tilde{\boldsymbol{s}} \lambda} \big)^2 
+ \left( t_i \, \Theta (\mp s_x)  + \Theta (\pm s_x) \right) \big( \boldsymbol{D}^{(i)}_{12} \cdot  \boldsymbol{e}_{\boldsymbol{s} \lambda} \big)^2 \right] \, ,
\end{eqnarray}
respectively. Which signs apply depends on whether $i$ equals $a$ or $b$. Moreover, introducing the notation $\boldsymbol{D}^{(i)}_{12}  = \| \boldsymbol{D}_{12} \| \, ( d^{(i)}_1, d^{(i)}_2, d^{(i)}_3)^{\rm T}$ with $|d^{(i)}_1|^{2} + |d^{(i)}_{2}|^{2} + |d^{(i)}_3|^{2} = 1$, one can now show that the above $\Gamma_{\rm mir}^{(ii)} $ both equal the free-space decay rate of an atom with dipole moment $\boldsymbol{D}^{(i)}_{12} = \boldsymbol{D}_{12}$,
\begin{eqnarray} \label{B15}
\Gamma_{\rm mir}^{(aa)}  = \Gamma_{\rm mir}^{(bb)} &=& \Gamma_{\rm free} ~~ {\rm with} ~~
\Gamma_{\rm free} = {e^{2} \omega_{0}^{3} \,  \| \boldsymbol{D}_{12} \|^{2} \over 3 \pi \hbar \varepsilon c^3} 
\end{eqnarray}
since $r_i + t_i = 1$. As we shall see below, this result does not mean that photons are emitted at their free-space rate $\Gamma_{\rm free}$, if initially only one of the two atoms is excited.

The two remaining constants $\Gamma_{\rm mir}^{(ab)} $ and $\Gamma_{\rm mir}^{(ba)}$ in Eq.~(\ref{L}) can be derived analogously. Since we are only interested in the case where the distance of each atom from the mirror interface and the distance between atom $a$ and atom $b$ are much larger than the wavelength of the emitted light, we can safely ignore terms describing direct interactions between both atoms and between an atom and its own mirror image. These are known to be relatively short-range. However, terms describing interactions between an atom and the mirror image of the atom on the opposite side must be kept, when ${\boldsymbol r}_a$ and $\tilde {\boldsymbol r}_b$ are relatively close. Using Eqs.~(\ref{eq:observables}) and (\ref{B9}) we therefore find that 
\begin{eqnarray} \label{B10x}
\gamma_{\rm mir}^{(ab)} (t,t')
&=& {16 \pi^3 \varepsilon \over \hbar c} \sum_{\lambda=1,2} \int {\rm d} \boldsymbol{s} \, \Theta(s_x) \bigg[ \, t_a(\tilde {\boldsymbol s}) r_b({\boldsymbol s}) \, 
\boldsymbol{D}^{(b)}_{12} \cdot \langle 0_{\rm F}| \boldsymbol{\cal E}_{\tilde {\boldsymbol s} \lambda} (\tilde {\boldsymbol r}_b + \tilde {\boldsymbol s} c t')^\dagger 
~ \boldsymbol{D}^{(a)}_{12} \cdot \boldsymbol{\cal E}_{\tilde {\boldsymbol s} \lambda} (\boldsymbol{r}_a + \tilde {\boldsymbol s} c t) |0_{\rm F} \rangle \notag \\
&& + \int {\rm d} \boldsymbol{s}' \, \Theta(-s_x') \, r_a({\boldsymbol s}') t_b({\boldsymbol s}) \, 
\boldsymbol{D}^{(b)}_{12} \cdot \langle 0_{\rm F}| \boldsymbol{\cal E}_{{\boldsymbol s} \lambda} ({\boldsymbol r}_b + {\boldsymbol s} c t')^\dagger 
~ \boldsymbol{D}^{(a)}_{12} \cdot \boldsymbol{\cal E}_{{\boldsymbol S}({\boldsymbol s}') \lambda} (\boldsymbol{R}_a ({\boldsymbol s}',t)) |0_{\rm F} \rangle \, \bigg]  
\end{eqnarray}
and $\gamma_{\rm mir}^{(ba)} (t,t') = \gamma_{\rm mir}^{(ab)} (t,t')^*$ to a very good approximation. If the scattering operator ${\boldsymbol S}$ scrambles the wave vectors of light originating from atom $a$ more or less randomly upon reflection, the second term in this equation becomes negligible. However, the first term in the above equation does not average away and is in general non-zero. For simplicity, let us assume that ${\boldsymbol r}_a$ and ${\boldsymbol r}_b$ have the same $y$ and $z$ coordinates. In this case
\begin{eqnarray} \label{B17b}
{\rm e}^{{\rm i} \boldsymbol{k} \cdot (\boldsymbol{r}_a - \tilde{\boldsymbol{r}}_b)} = {\rm e}^{{\rm i} k \cos \vartheta (x_a + x_b)} \, .
\end{eqnarray}
Replacing $t_a(\tilde {\boldsymbol s})$ and $r_b({\boldsymbol s})$ by their average values in Eq.~(\ref{oma100}), proceeding as described in the previous subsection and using again Eq.~(\ref{B13}), leads us to
\begin{eqnarray} \label{B18}
\Gamma_{\rm mir}^{(ab)} = {3  t_a r_b \Gamma_{\rm free} \over  8 \pi} \sum_{\lambda=1,2} 
\int_{\pi/2}^{\pi} {\rm d}\vartheta \, {\rm e}^{-{\rm i} \xi \cos \vartheta} \, \sin \vartheta
\int_0^{2 \pi} {\rm d}\varphi \, \left( \hat{\boldsymbol{D}}^{(a)}_{12} \cdot \boldsymbol{e}_{\boldsymbol{s} \lambda} \right) \left( \hat{\boldsymbol{D}}^{(b)}_{12} \cdot \boldsymbol{e}_{\boldsymbol{s} \lambda} \right) \, .
\end{eqnarray}
The hat symbols indicate that the vectors $\boldsymbol{D}^{(i)}_{12}$ have been normalised, the polarisation vectors $\boldsymbol{e}_{\boldsymbol{s} \lambda}$ can be found in Eq.~(\ref{vectors}), and $\xi = k_0(x_a + x_b)$ with $k_0 = \omega_0/c$ is a relative effective distance. Performing the $\varphi$ integration and substituting $u = - \cos\vartheta$ yields
\begin{eqnarray} \label{B19}
\Gamma_{\rm mir}^{(ab)} = {3 t_a r_b  \Gamma_{\rm free} \over 16} \int_0^1 {\rm d}u \, {\rm e}^{{\rm i} \xi u}  \left[  2 d_1^{(a)} d_1^{(b)} \left( 1 - u^2 \right) + \left( d_2^{(a)} d_2^{(b)} + d_3^{(a)} d_3^{(b)} \right) \left( 1 + u^2 \right) \right] . 
\end{eqnarray}
Performing the final integration, the above constant simplifies to
\begin{eqnarray} \label{Bfin}
\Gamma_{\rm mir}^{(ab)} = {3 t_a r_b  \Gamma_{\rm free} \over 16} \left[ \sum_{i=2}^3 d_i^{(a)} d_i^{(b)} \left( 2 {\rm e}^{{\rm i} \xi} \left( {1 \over {\rm i} \xi} + {1 \over \xi^2} - {1 \over {\rm i} \xi^3} \right) - {1 \over {\rm i} \xi} + {2 \over {\rm i} \xi^3} \right) - 2 d_1^{(a)}  d_1^{(b)} \left( 2{\rm e}^{{\rm i} \xi} \left( {1 \over \xi^2} - {1 \over {\rm i} \xi^3} \right) + {1 \over {\rm i} \xi} + {2 \over {\rm i} \xi^3} \right) \right] ~~
\end{eqnarray}
which coincides with Eq.~(\ref{B17}) in the main text. Analogously, one can show that $\Gamma_{\rm mir}^{(ba)} =\Gamma_{\rm mir}^{(ab)*}$. If the $y$ and the $z$ coordinates of the position of atom $a$ and atom $b$ are not the same, additional terms have to be taken into account in the above derivation. However, our physical intuition tells us that the remote interaction between atom $a$ and atom $b$ depends also in this case only on the relative effective distance $\xi$ and not on the actual distance of the atoms.

\subsection{The reset operator ${\cal L} (\rho_{\rm A})$}

For completeness, we now also calculate the state ${\cal L} (\rho_{\rm A})$ of the atoms in case of an emission. Substituting Eq.~(\ref{eq:Hdef3}) into Eq.~(\ref{BB5}) and introducing the variables $\tau = \Delta t-t$ and $\tau' = \Delta t-t'$ yields
\begin{eqnarray} \label{L2}
\mathcal{L} (\rho_{\rm A}) = \sum_{i,j = a,b} \widetilde \Gamma_{\rm mir}^{(ij)} \, \sigma_{i}^{-} \rho_{\rm A} \sigma_{j}^{+} 
\end{eqnarray}
 with $\widetilde \Gamma_{\rm mir}^{(ij)}$ given by
\begin{eqnarray} \label{B8xxx}
\widetilde \Gamma_{\rm mir}^{(ij)} &=& {e^2 \over \hbar^2 \Delta t} \int_0^{\Delta t} {\rm d} \tau \int_0^{\Delta t} {\rm d} \tau' \, {\rm e}^{{\rm i} \omega_0 (\tau - \tau')} \, \boldsymbol{D}^{(j)}_{12} \cdot \langle 0_{\rm F}| \boldsymbol{\cal E} (\boldsymbol{r}_j) \, U^\dagger_{\rm F}(\tau',0) ~ \boldsymbol{D}^{(i)}_{12} \cdot U_{\rm F}(\tau,0) \boldsymbol{\cal E}(\boldsymbol{r}_i)^\dagger \, |0_{\rm F} \rangle \, .
\end{eqnarray}
These constants have many similarities with the constants $\Gamma_{\rm mir}^{(ij)}$ in Eqs.~(\ref{B8}). The only differences are a missing factor 2 and a different upper limit on the second time integral. Proceeding as in the previous subsection, we find that evaluating Eq.~(\ref{B8xxx}) now leads to time integrals of the form
\begin{eqnarray} \label{B10xxx}
\int_0^{\Delta t} {\rm d} \tau \int_0^{\Delta t} {\rm d} \tau' \, {\rm e}^{{\rm i} \omega (\tau - \tau')} 
= 2 {\rm Re} \left( \int_0^{\Delta t} {\rm d} \tau \int_0^\tau {\rm d} \tau' \, {\rm e}^{{\rm i} \omega (\tau - \tau')} \right)
\end{eqnarray}
\end{widetext} 
with $\omega = \omega_0-ck$. Hence all the constants $\widetilde \Gamma_{\rm mir}^{(ij)}$ are real and $\widetilde \Gamma_{\rm mir}^{(ij)} = {\rm Re} \left( \Gamma_{\rm mir}^{(ij)} \right) $ which yields the reset operator $\mathcal{L} (\rho_{\rm A})$ in Eq.~(\ref{L}). \\[0.5cm]
\noindent {\bf Funding.} 
We acknowledge financial support from the Oxford Quantum Technology Hub NQIT (grant number EP/M013243/1) and the EPSRC (2115757). \\[0.2cm]
\noindent {\bf Competing Interest:} 
The authors have no relevant financial or non-financial interests to disclose. \\[0.2cm]
\noindent {\bf Author Contributions:} 
Conceptualisation, G.J. and A.B.; methodology, N.F.-W., B.D. and A.B.; formal analysis, N.F.-W., B.D. and A.B.; writing and original draft preparation, N.F.-W., B.D. and A.B.; writing and review and editing, T.M. and G.J.; visualization, B.D. and A.B.; supervision, A.B.; funding acquisition, N.F.-W., G.J. and A.B. All authors have read and agreed to the published version of the manuscript. \\[0.2cm]
\noindent{\bf Data availability:} 
This publication is theoretical work that does not require supporting research data. \\


\end{document}